\documentclass{raa}

\usepackage{graphicx,times,xcolor}         
\usepackage{natbib}
\usepackage{ulem}
\usepackage{amssymb,amsmath}

\usepackage[pagebackref=true]{hyperref}

\newcommand{\g}{$\gamma$-ray}
\newcommand{\lat}{\textit{Fermi}-LAT}

\begin{document}

  \title{Diffuse Gamma-ray Emission Around 4FGL~J1626.0$-$4917
}

   \volnopage{Vol.0 (20xx) No.0, 000--000}      
   \setcounter{page}{1}          

   \author{Ziwei Ou 
      \inst{1}
   \and Jie Wang
      \inst{2,3}
   }


   \institute{Tsung-Dao Lee Institute, Shanghai Jiao Tong University,
             Shanghai 201210, China; {\it ziwei@sjtu.edu.cn}\\
        \and
             Xinjiang Astronomical Observatory, Chinese Academy of Sciences, Urumqi, 830011, China; {\it wangjie@xao.ac.cn}\\
        \and
            University of Chinese Academy of Sciences, Beijing 100049, China\\
\vs\no
   {\small Received 2025 September day; accepted 20xx month day}}

\abstract{Extended gamma-ray sources provide significant information about particle propagation. 4FGL~J1626.0$-$4917 was labeled as an unassociated source in 4FGL catalog without known counterparts at other wavelengths. We report an analysis on 4FGL~J1626.0$-$4917 with 17 years \textit{Fermi} Large Area Telescope data and archival \textit{Chandra} X-ray Observatory data. We find extended GeV emission around this source, which can be modelled by a Gaussian disk of $0.28^{\circ}$ radius with a significance of the extension of 7.2$\sigma$. The gamma-ray spectrum of 4FGL~J1626.0$-$4917 has a photon index of $2.73 \pm 0.12$. The gas content, including molecular, neutral and ionized gas, was investigated and the potential hadronic origin is discussed. The derived gas mass around 4FGL~J1626.0$-$4917 is $\sim 1.1 \times 10^3 M_{\odot}$. The diffuse GeV gamma-ray emission may likely originate from the interaction between accelerated protons in 4FGL~J1626.0$-$4917 and the target proton in surrounding gas, although the leptonic process cannot be ruled out. The X-ray spectral analysis was performed, which reveal a point source inside 4FGL~J1626.0$-$4917. We investigate potential counterparts, including the stellar cluster NGC~6134 and the supernova remnant G335.2$+$0.1. Our results highlight the complexity of unidentified extended gamma-ray sources and the need for further observations.
}

   \authorrunning{Ou \& Wang}            
   \titlerunning{4FGL~J1626.0$-$4917}  

   \maketitle

%
%
\section{Introduction}         \label{sect:intro}

During the 17 years for its operation, \lat\ has found numerous high-energy \g\ sources \citep{Abdo2010b,Nolan2012,Ajello2017,Abdollahi2020}. For about one-third sources in \lat\ Telescope Fourth Source Catalog (4FGL) catalog, no counterparts have been found at other wavelengths. Research of \lat\ unidentified sources focus on two aspects. Firstly, machine learning methods are employed to classify unidentified sources based on prior knowledge \citep[e.g.][]{SazParkinson2016,Ding2023,Ou2025}. Secondly, new multi-wavelength programs are conducted to determine the nature of these sources \citep[e.g.][]{Acero2013,Paiano2017,Hui2020,Ulgiati2025}. The \g\ spectral and morphological information would assist us in understanding the nature of \lat\ sources. Observations from the \lat\ on supernova remnants (SNRs) \citep{Araya2024}, pulsar wind nebulae (PWNe) \citep{Tibaldo2018,Principe2020} and pulsar halos \citep{DiMauro2019,Abdollahi2024} found extended \g\ emissions around them. Such extended \g\ sources are found at low Galactic-latitude \citep{Abdollahi2024} and high Galactic-latitude \citep{Ackermann2018}. Furthermore, \cite{Xiang2024} found faint extended sources in \lat\ data which had not been reported by \lat\ . Both population study \citep{Sato2021,Orlando2022} and individual source study \citep[e.g.][]{Xin2019,Shi2025} may help us to understand particle transportation around compact objects.

4FGL~J1626.0$-$4917 (l=334.69, b=-0.11) is an unidentified source reported by \lat\ Collaboration in 4FGL catalog. It locates near SNR~G335.2+0.1 and stellar cluster NGC~6134, which makes the nature of \g\ emissions from this source be valuable. Moreover, an unidentified TeV source HESS~J1626-490 was found in this region \citep{Abdalla2018}. In this paper, we report the \g\ and X-ray data analysis and discuss potential origin of the \g\ emissions. We performed a detailed study on \g\ emissions around 4FGL~J1626.0$-$4917 in Section~\ref{sec:fermi-data}. The morphology, extension and spectral index of 4FGL~J1626.0$-$4917 are given. To investigate the potential hadronic signal from a \g\ source, we study the gas distribution, including H$_2$, HI, and HII in Section~\ref{sec:gas}. To further obtain more information about 4FGL~J1626.0$-$4917, we performed X-ray spectral analysis with data from \textit{Chandra} X-ray Observatory, which are shown in Section~\ref{sec:x-ray}. In Section~\ref{sec:discus}, we discuss the potential origin of 4FGL~J1626.0$-$4917, including stellar cluster (Section~\ref{subsec:cluster}) and SNR (Section~\ref{subsec:snr}). Finally, we summary our main conclusion in Section~\ref{sec:conclu}.

\section{\lat\ Data Analysis} \label{sec:fermi-data} 

The \lat\ data utilized in this study spans from August 8 2008 to May 19 2025. We selected all events within a region-of-interest (ROI) of radius $15^{\circ}$ for 4FGL~J1626.0$-$4917. Data energy range from 300 MeV to 300 GeV were adopted in this analysis. The following criterion were adopted for the data: \texttt{zmax==90, evclass==128, evtype==3, DATA\_QUAL>0, LAT\_CONFIG==1}. Its coordinates are R.A.=246.5083 and Dec.=-49.2838, given by the DR4 of 4FGL catalog. The analysis was performed using the Fermi Science Tools \textit{v11r5p3}  and Fermipy \textit{1.2.0}  packages. For the diffuse background components, we adopt the Galactic diffuse emission model gll\_iem\_v07 and isotropic extragalactic emission model P8\_R3\_V3.

\subsection{Spatial Analysis} \label{subsec:spatial}

We initiate our analysis by searching extension of  4FGL~J1626.0$-$4917. Extension is one of the significant factor for help us to unveil the nature of a \g\ source. A source extension analysis is executed by performing a likelihood ratio test with respect to the no-extension (i.e. point source) with a best-fit model for extension. 

\begin{equation}
    TS_{\rm ext} = -2 \left( \mathrm{ln}\left(L_{\rm PS}\right) -  \mathrm{ln}\left(L_{\rm ext}\right) \right)
\end{equation}

Both a radial Guassian model and a radial disk are used for obtained $TS_{\rm ext}$. The best-fit extension values are determined by performing a likelihood profile scan over the 68\% containment and fitting for the extension which maximizes the model. For 4FGL~J1626.0$-$4917, we got a extension of $0.28 \pm 0.03 ^{\circ}$ and $0.26 \pm 0.02 ^{\circ}$ for radial Gaussian and disk models. The corresponding TS$_{\rm ext}$ are 52 and 50, respectively. In addition, the TS value of detection is given by TS = 273. Therefore, we adopt $0.28^{\circ}$ as extension in the following analysis.

\begin{figure}
    \centering
    \includegraphics[width=0.75\textwidth]{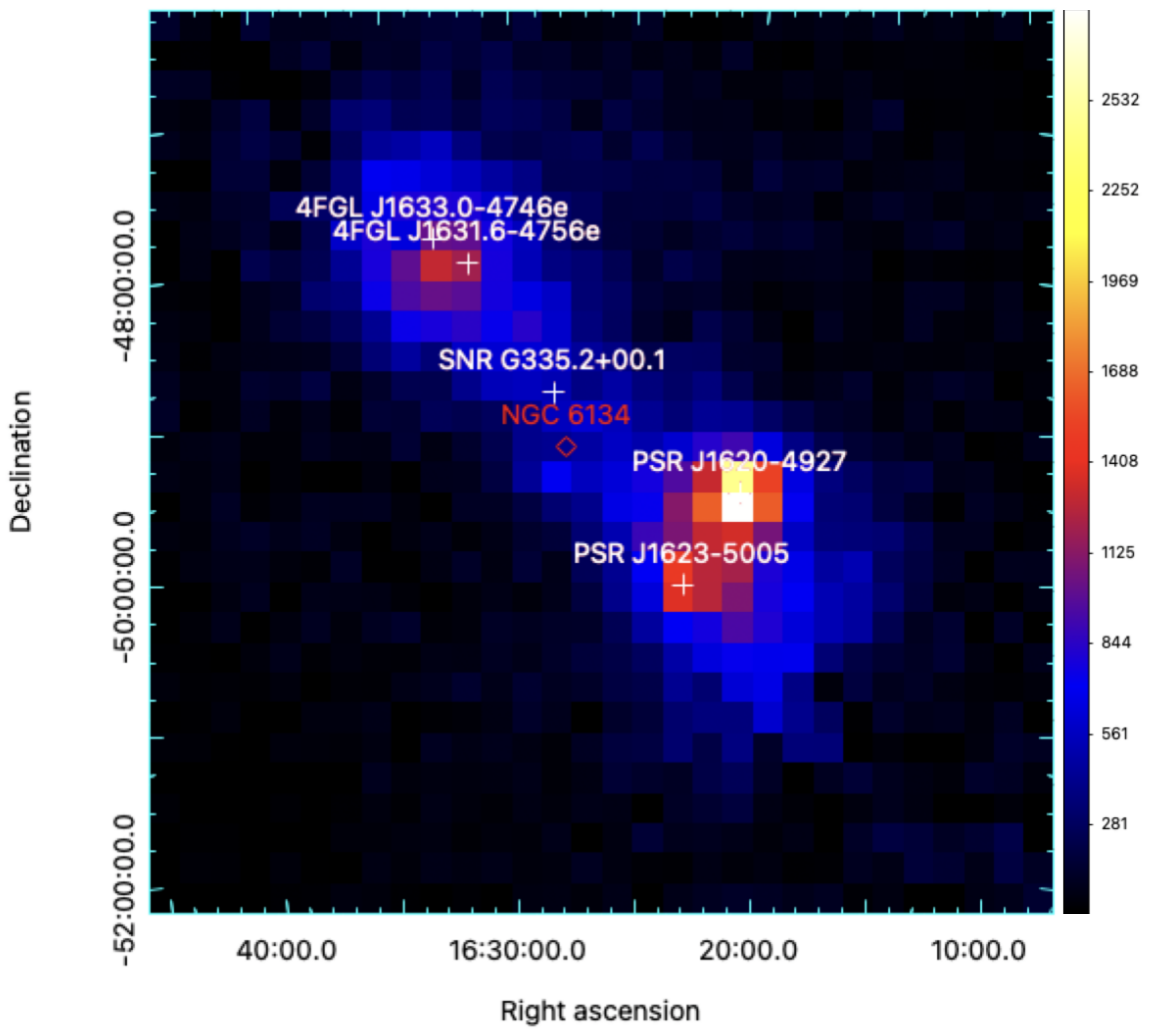}
    \caption{\g\ counts map (300 MeV to 300 GeV) in equatorial coordinate (R.A., Dec.). The sources out of $2^{\circ}$ radius, the Galactic diffuse emission and the isotropic radiation are removed. The counterparts of identified 4FGL sources are labeled. NGC~6134 is shown also.}
    \label{fig:known}
\end{figure}

\begin{figure}
    \centering
    \includegraphics[width=0.75\textwidth]{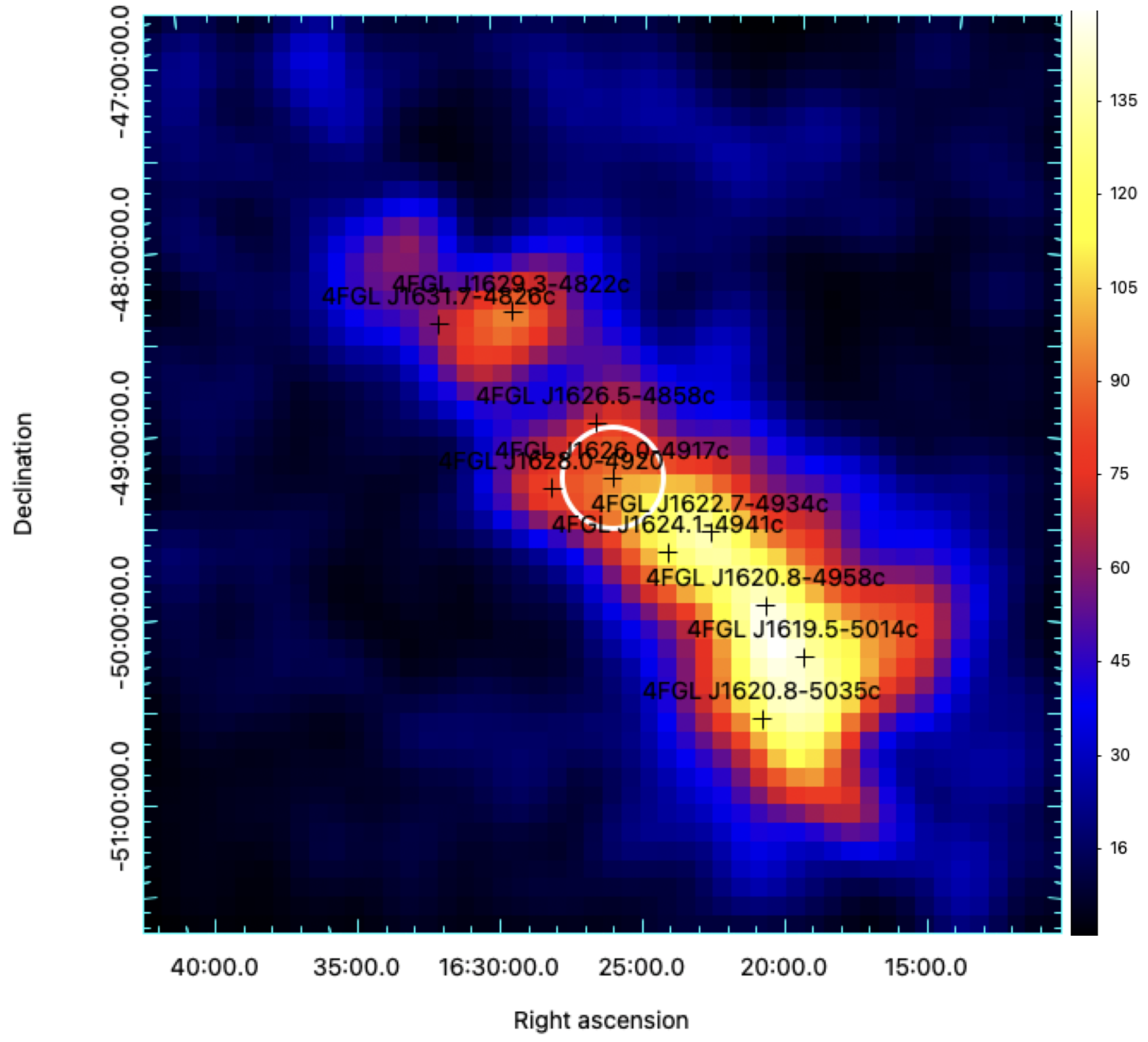}
    \caption{\g\ counts map (300 MeV to 300 GeV) in equatorial coordinate (R.A., Dec.). The 4FGL sources with unknown nature are labeled. The white circle represents the extension of 4FGL~J1626.0$-$4917, which is 0.28$^{\circ}$ as given in Section~\ref{subsec:spatial}}
    \label{fig:unknown}
\end{figure}

We show the sources only in the radius of $2^{\circ}$ with known sources labeled in Figure~\ref{fig:known}. A stellar cluster NGC~6134 was found in the center of ROI. Also, the Galactic diffuse emission and the isotropic radiation are subtracted. Since we are planing to explore the extended \g\ emissions around 4FGL~J1626.0$-$4917, we then subtract the 4FGL sources associated with known objects. These include PSR~J1620$-$4927, PSR~J1623$-$5005 and SNR~G335.2$+$00.1. In addition, 4FGL~J1633.0$-$4746e and 4FGL~J1631.6$-$4756e, which are associated with HESS~J1632$-$478 (PWN) are excluded also \citep{Balbo2010}. The counts map with labeled 4FGL unassociated sources is shown in Figure~\ref{fig:unknown}.

\begin{figure}
    \centering
    \includegraphics[width=0.75\linewidth]{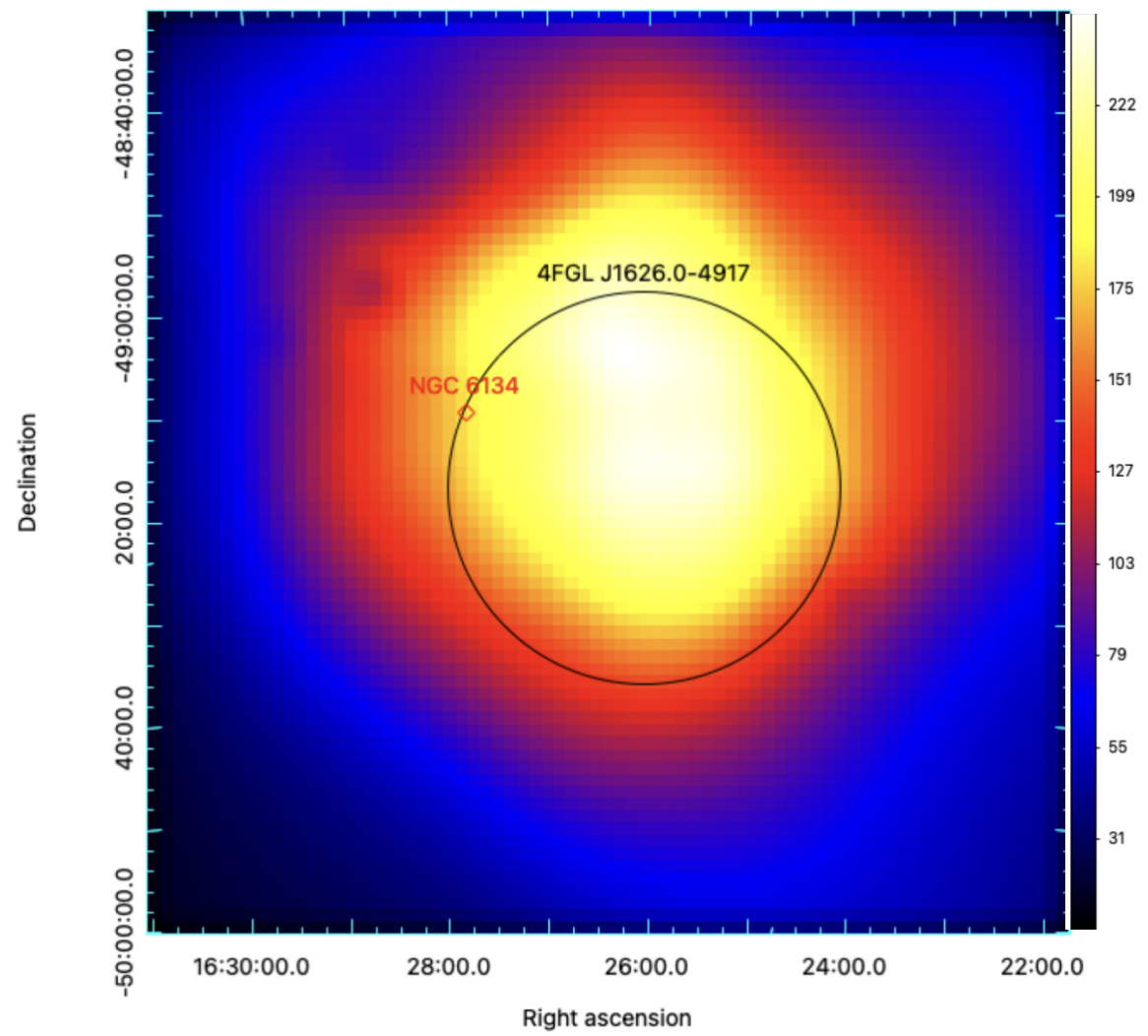}
    \caption{The $1.5^{\circ} \times 1.5^{\circ}$ \g\ TS map (300 MeV to 300 GeV) around 4FGL~J1626.0$-$4917.}
    \label{fig:ts-map}
\end{figure}

There are three pulsars in the $0.5^{\circ}$ around of 4FGL~J1626.0$-$4917 \footnote{https://www.atnf.csiro.au/research/pulsar/psrcat/}: PSR~J1623$-$4931, PSR~J1625$-$4904 and PSR~J1625$-$4913. None of them are detected by 3PC \citep{Smith2023}. In addition, PSR~J1625$-$4904 and PSR~J1625$-$4913 have $\dot{E}$ lower than $10^{34}\, \rm erg/s$ while PSR~J1625$-$4904 does not have measured $\dot{E}$. These indicate that the extended \g\ emission of 4FGL~J1626.0$-$4917 disfavored a halo origin from one of these pulsars. 4FGL~J1626.5$-$4858c and 4FGL~J1628.0$-$4920 locates near the position of 4FGL~J1626.0$-$4917 with offset as $18.7^{\prime}$ and $19.9^{\prime}$. However, the nature of these two objects are unclear. We removed them from the total source model. The TS map was provided in Figure~\ref{fig:ts-map}. It show that NGC~6134 inside the extension of 4FGL~J1626.0$-$4917.

\subsection{Spectral Analysis}

The spectral shape aids us in judging the nature of \g\ sources. To validate the spectral model, we test for spectral curvature that reveals deviations from a PL spectrum ($\frac{dN}{dE} = N_0 (\frac{E}{E_0})^{-\Gamma}$) for each source via likelihood ratio test. For PL spectrum, $N_0$ is normalization, $E_0$ is pivot energy, and $\Gamma$ is photon index. For the LogParabola (LP) model, it provides: $TS_{\rm LP} = -2 \left( \mathrm{ln}\left(L_{\rm PL}\right) - \mathrm{ln}\left(L_{\rm LP}\right) \right)$

LP model is denfined as:
\begin{equation}
    \frac{dN}{dE} = N_0 \left( \frac{E}{E_b} \right)^{-(\alpha + \beta\ln(E/E_b))}
\end{equation}
where $\alpha$ is photon index, $\beta$ is the second index, and $E_b$ is the scale parameter.

Similarly, for PLSuperExpCutoff (PLSC) model, it gives: $TS_{\rm PLSC} = -2 \left( \mathrm{ln}\left(L_{\rm PL}\right) -  \mathrm{ln}\left(L_{\rm PLSC}\right) \right)$.

PLSC model is defined as:
\begin{equation}
    \frac{dN}{dE} = N_0 \left( \frac{E}{E_0} \right)^{\gamma_0-\frac{d}{2}\ln\frac{E}{E_0} - \frac{db}{6} \ln^2\frac{E}{E_0} - \frac{db^2}{24} \ln^3\frac{E}{E_0}} 
\end{equation}
where $\gamma_0$ is the photon index, $d$ is the parameter describe the shape of he exponential cutoff, $b$ is the local curvature parameter.

\begin{figure}
    \centering
    \includegraphics[width=0.75\textwidth]{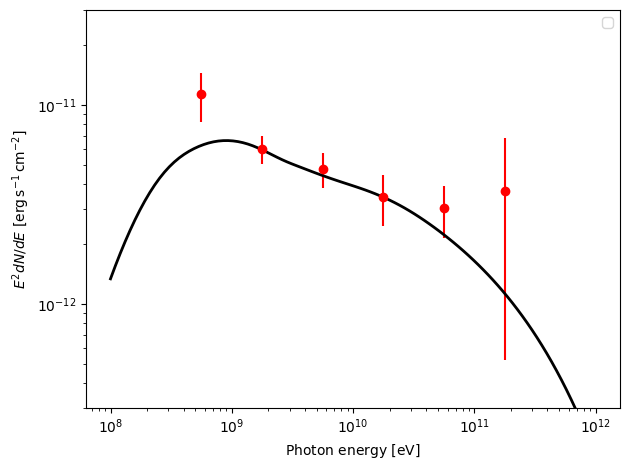}
    \caption{\g\ spectrum of 4FGL~J1626.0$-$4917 with \lat\ data. The hadronic model is shown also, which will be discussed in Section~\ref{sec:gas}.}
    \label{fig:sed}
\end{figure}

Once TS$_{\rm LP}$ or TS$_{\rm PLSC}$ is larger than 25, we perform a spectral fit for LP and PLSC models, otherwise, the PL model is adopted. The curvature tests show that TS$_{\rm LP}$ = 0 and TS$_{\rm PLSC}$ = 0. Therefore, we use PL as spectral model to fit the data of 4FGL~J1626.0$-$4917, which obtain spectral index $\Gamma = 2.73 \pm 0.12$. The corresponding spectrum is shown in Figure~\ref{fig:sed}.

\section{Gas Content around 4FGL~J1626.0$-$4917}\label{sec:gas}

\begin{figure*}
    \centering
    \includegraphics[width=0.3\textwidth]{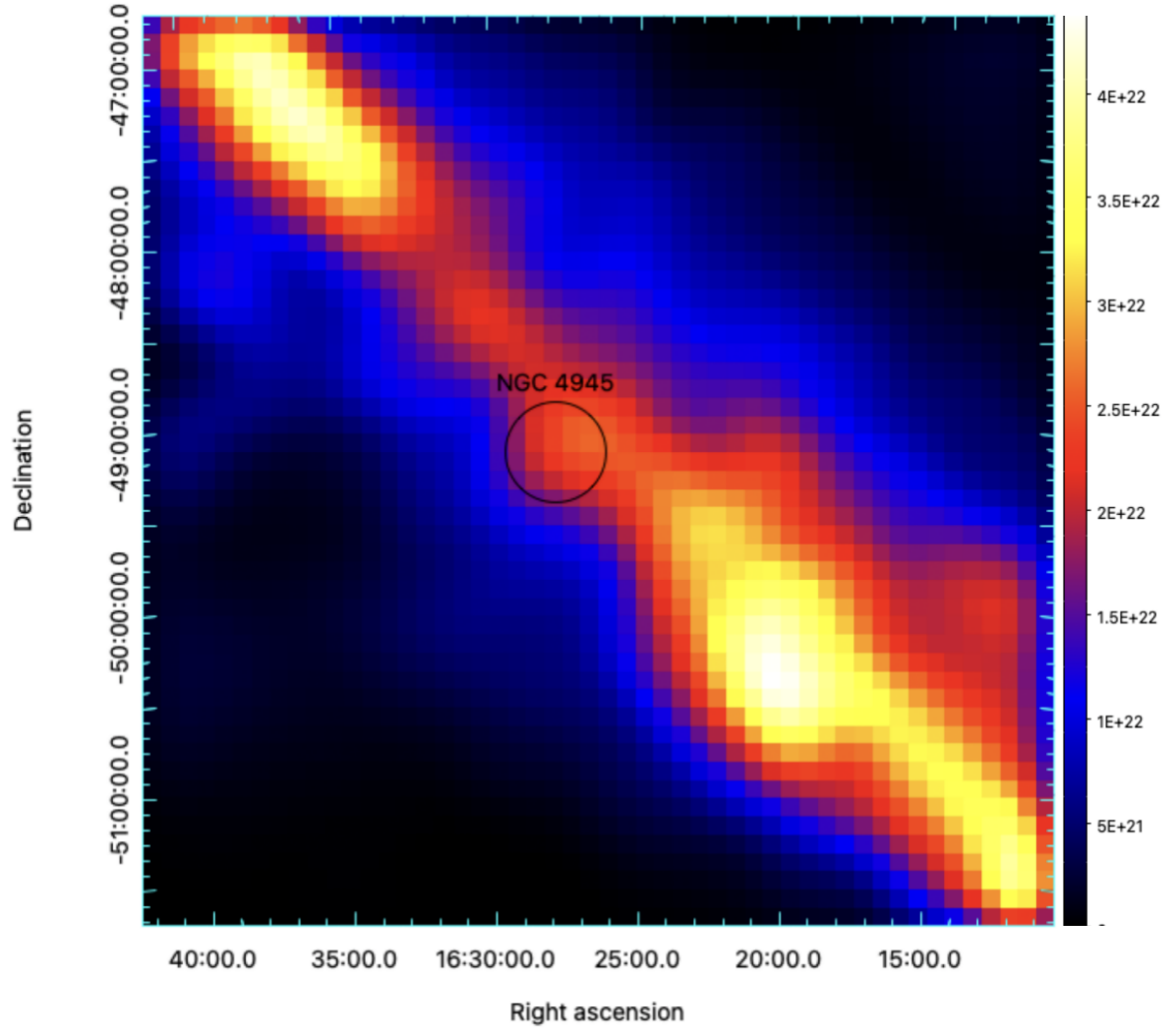}
    \includegraphics[width=0.3\textwidth]{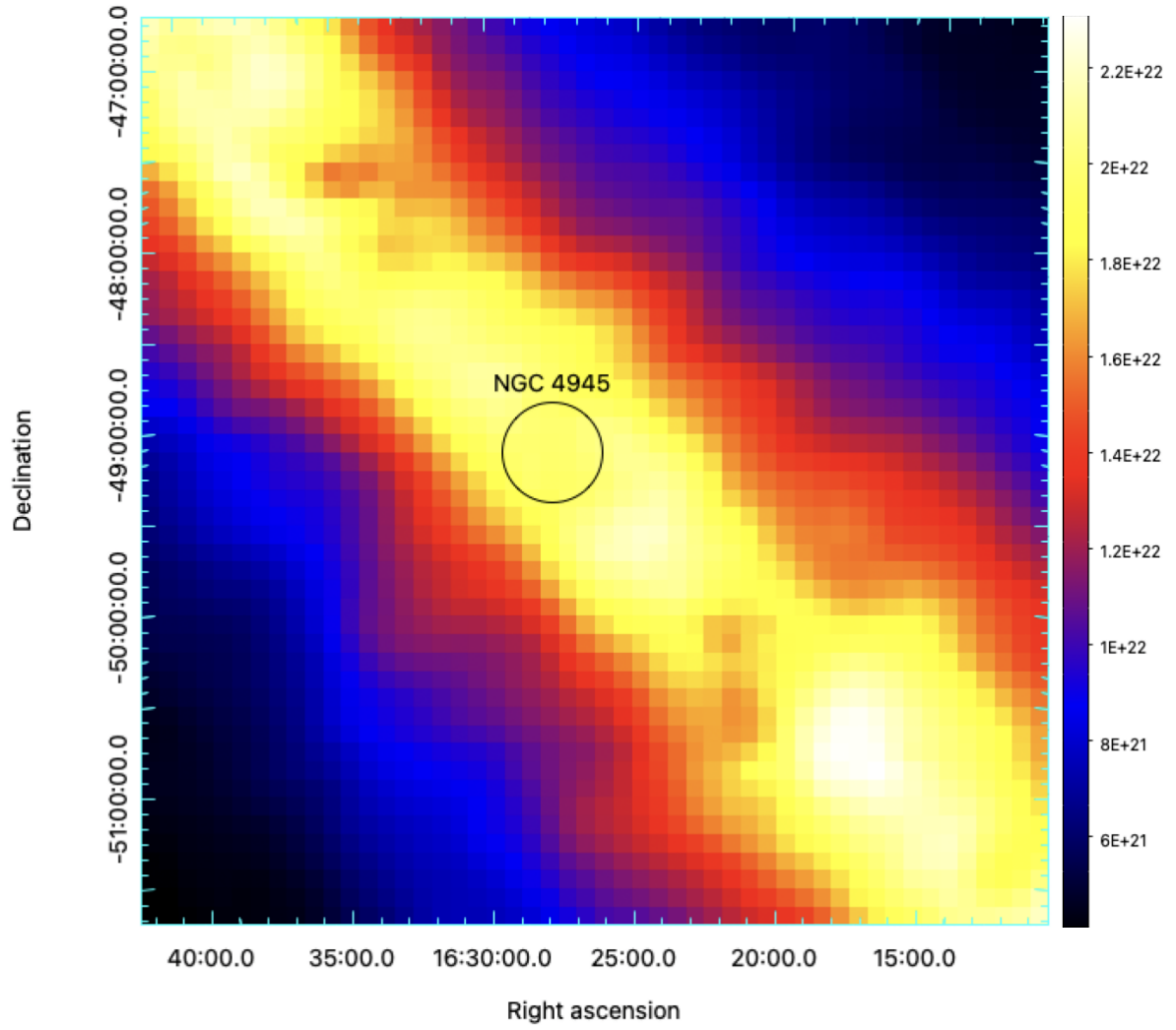}
    \includegraphics[width=0.3\textwidth]{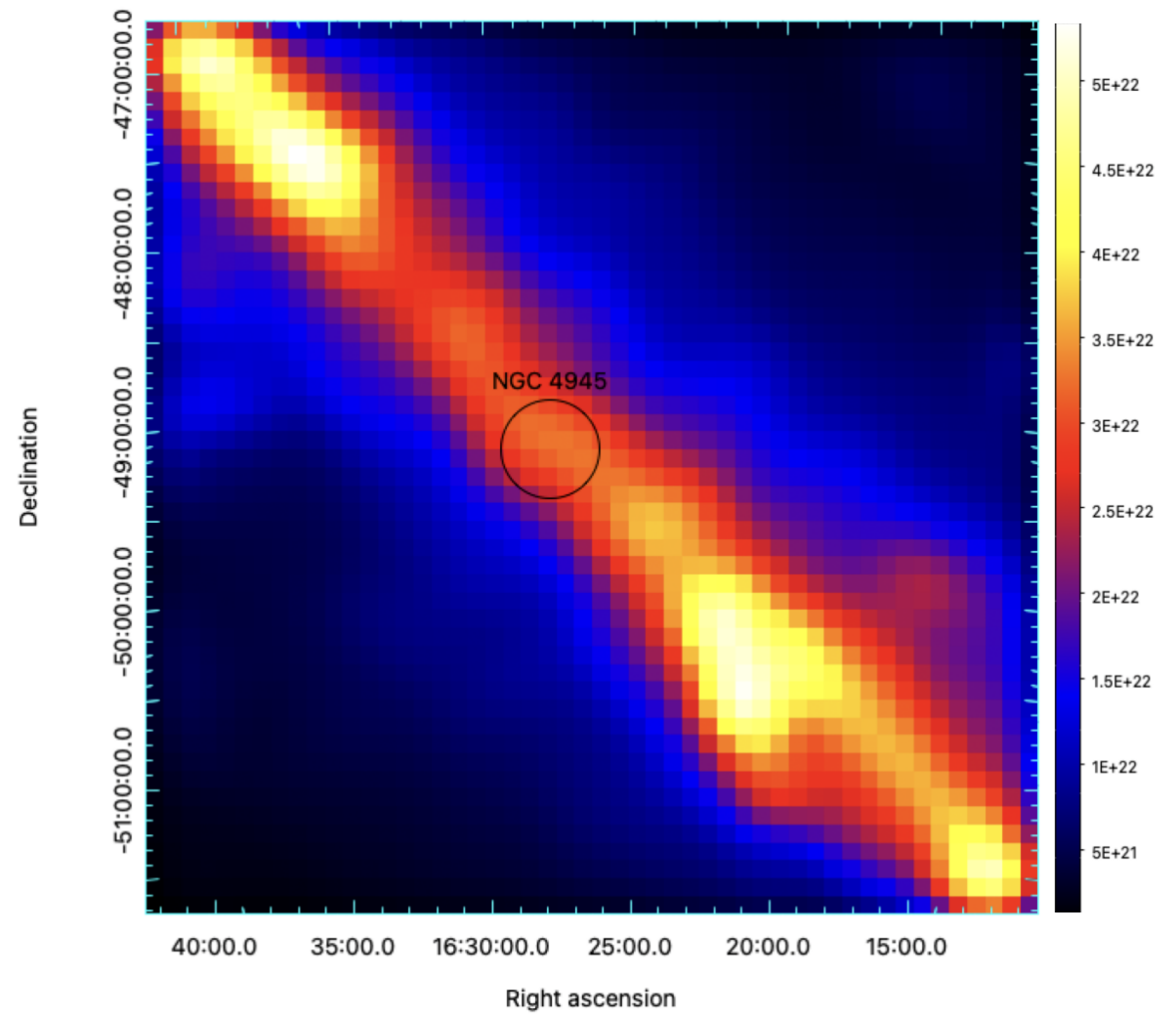}
    \caption{Gas column densities (in unit of $\rm cm^{-2}$) with different gas phase in equatorial coordinate. The left panel gives the $\rm H_2$ column density derived from CO data. The middle panel provides the map of H I column density derived from a 21-cm all-sky survey. The right panel shows the H II column density derived from \textit{Planck} free-free (353 GHz) map assuming an effective density of electron $n_e = 10\ \rm cm^{-3}$.} 
    \label{fig:gas}
\end{figure*}

For a \g\ source, one of the key questions is whether the \g\ emissions are dominated by leptonic or hadronic processes. This is important for both stellar clusters and SNRs. The most significant hadronic process is pp interaction, which originates from the interaction between protons from the \g\ sources and target protons in surrounding medium \citep{Kafexhiu2014}. Therefore, we investigate the H$_2$, the neutral atomic hydrogen HI, and the HII gas distribution in the vicinity of 4FGL~J1626.0$-$4917. We follow the procedure from \cite{Ge2022}. To investigate the H$_2$ distribution toward 4FGL~J1626.0$-$4917, we use CO data from \cite{Dame2001}. The $^{13}$CO (J = 1-0) line profile of the molecular cloud may reflect the kinematic activities of the gas distribution. The velocity distribution of the CO is considered to be -27 to -18 $\rm km\ s^{-1}$, corresponding to a mean kinematic distance of 1.8 kpc \citep{Eger2011}. This velocity is used for atomic gas (HI) also. The gas distribution in this region can be found in Figure~\ref{fig:gas} (left).

The HI data are obtained from the HI 4$\pi$ survey (HI4PI) \citep{HI4PI2016}. This survey aims to obtain all-sky Galactic HI data with 21 cm hydrogen lines.

\begin{equation}
   N_{\rm HI} = -1.38 \times 10^{18}\ T_{\rm s} \int d\nu \ln (1- \frac{T_{\rm B}}{T_{\rm s} - T_{\rm bg}})
\end{equation}

We adopted $T_{\rm bg} = 2.66\ \rm K$ as the brightness temperature of the cosmic microwave background. $T_{\rm B}$ is the brightness temperature of the HI emission. For $T_{\rm B} > T_{\rm s} - 5\ \rm K$, $T_{\rm B}$ can be addressed as $T_{\rm s} - 5\ \rm K$, and $T_{\rm s}$ is adopted to 150 K. The integral velocity range is the same as that of CO gas. The HI column density map can be seen in Figure~\ref{fig:gas} (middle).

The H II column density can be derived from \textit{Planck} free-free map \citep{Planck2016}. The conversion factor from emission measure to free-free intensity can be found in \cite{Finkbeiner2003}. The column density from the intensity of the free-free emission can be calculated with 

\begin{equation} 
    N_{\rm H II} = 1.2 \times 10^{15} \ \rm cm^{-2} (\frac{T_e}{1 K})^{0.35} (\frac{\nu}{1 GHz})^{0.1} (\frac{n_e}{1 cm^{-3}})^{-1} \times \frac{I_{\nu}}{1 \rm Jy\ sr^{-1}}
\end{equation}

In this case, the frequency $\nu = 353\ \rm GHz$, electron temperature $T_e = 8000\ \rm K$, and the effective electron density $n_e = 10\ \rm cm^{-3}$ are adopted in the calculation. In addition, $I_{\nu} = 46.04\ \rm Jy\ sr^{-1}$ is adopted corresponding to $\nu = 353\ \rm GHz$ \citep{Finkbeiner2003}. The corresponding column density map can be found in Figure~\ref{fig:gas} (right).

The column density of H$_2$ can be defined by $N_{\rm H_2} = X_{\rm CO} \times W_{\rm CO}$. The conversion factor $X_{\rm CO}$ is set to $2 \times 10^{20}\ \rm cm^{-2}\ (K\ km\ s^{-1})$ \citep{Dame2001,Bolatto2013}. The intensity of the CO line $W_{\rm CO}$ (in unit of $\rm K\ km\ s^{-1}$) was calculated using velocity range of 40-55 $\rm km\ s^{-1}$. The total mass of the molecular cloud can be calculated with 

\begin{equation}
    M = \mu m_{\rm H} d^2 \Omega_{px} X_{\rm CO} \sum_{px} W_{\rm CO}
\end{equation}

where $\mu$ reflects the mean molecular weight, $m_{\rm H}$ is the mass of the H function, and $\Omega_{px}$ is the solid angle for each pixel. The spatial distribution of GeV \g\ emission within the region is spherical in geometry. The estimated mass and number density for three gas phase are shown in Table~\ref{tab:gas}.

Considering the hadronic process, we fit the \lat\ spectrum with Naima \citep{Zabalza2015}. Since the derived average number density of the target protons for this region is $N_{\rm H} = 3194.01\ \rm cm^{-3}$ has already in previous calculation (Table~\ref{tab:gas}), we fixed this value in our analysis. We adopted a power-law with exponential cutoff spectrum for the parent proton distribution $f_{\rm p}(E) = A_{\rm p} (E/E_0)^{-\alpha_{\rm p}}\exp (- (E / E_{\rm c})$. The best-fit parameters are: $\alpha_{\rm p} = 2.32 \pm 0.14$, and $E_{\rm c} = 8.91 \pm 0.82\ \rm TeV$. Considering the distance of 1.8 kpc \citep{Eger2011}, the total energy $W_{\rm p} = (5.45 \pm 0.23) \times 10^{47}\ \rm erg$ for the proton above 1 GeV was obtained.

\begin{table}[]
    \centering
    \begin{tabular}{c|c|c}
    \hline
        Tracer & Mass ($10^2\ \rm M_{\odot}$) & Number density ($\rm cm^{-3}$)\\\hline
        H$_2$ & 3.24 & 898.23 \\
        H I & 3.88 & 1076.83 \\
        H II & 4.39 & 1218.96 \\
        total & 11.52 & 3194.01 \\
        \hline
    \end{tabular}
    \caption{Gas mass and number density within the Gaussian disk with a radius of $0.28^\circ$.}
    \label{tab:gas}
\end{table}

\section{X-ray Data Analysis} \label{sec:x-ray}

We perform an X-ray study on 4FGL~J1626.0$-$4917 to check whether we can find potential counterpart for it. It was observed by Advanced CCD Imaging Spectrometer in \textit{Chandra} one time (ObsID: 13287, PI: Garmire) for 10.04 ks in June 2012. The count rate is around $33.53 \pm 1.83\ \rm cts/s$. We analyze the archive \textit{Chandra} data using \textit{Chandra} Interactive Analysis of Observation (CIAO) v4.13 and calibration files CALDB 4.9.6. The \texttt{chandra\_repro} tool was used to reprocess and reduce the data. Subsequently, the \textsc{wavdetect} tool from CIAO was employed to detect point sources in the region. This step identifies probable source pixels within the dataset by correlating it with wavelet functions repeatedly and generates a source list with information from each wavelet scale. Only one source with significance greater than $5\sigma$ was detected. After that, we used \texttt{specextract} tool to extract spectrum, and XSPEC v12.13.0 to fit the spectra.

\begin{figure}
    \centering
    \includegraphics[width=0.65\textwidth]{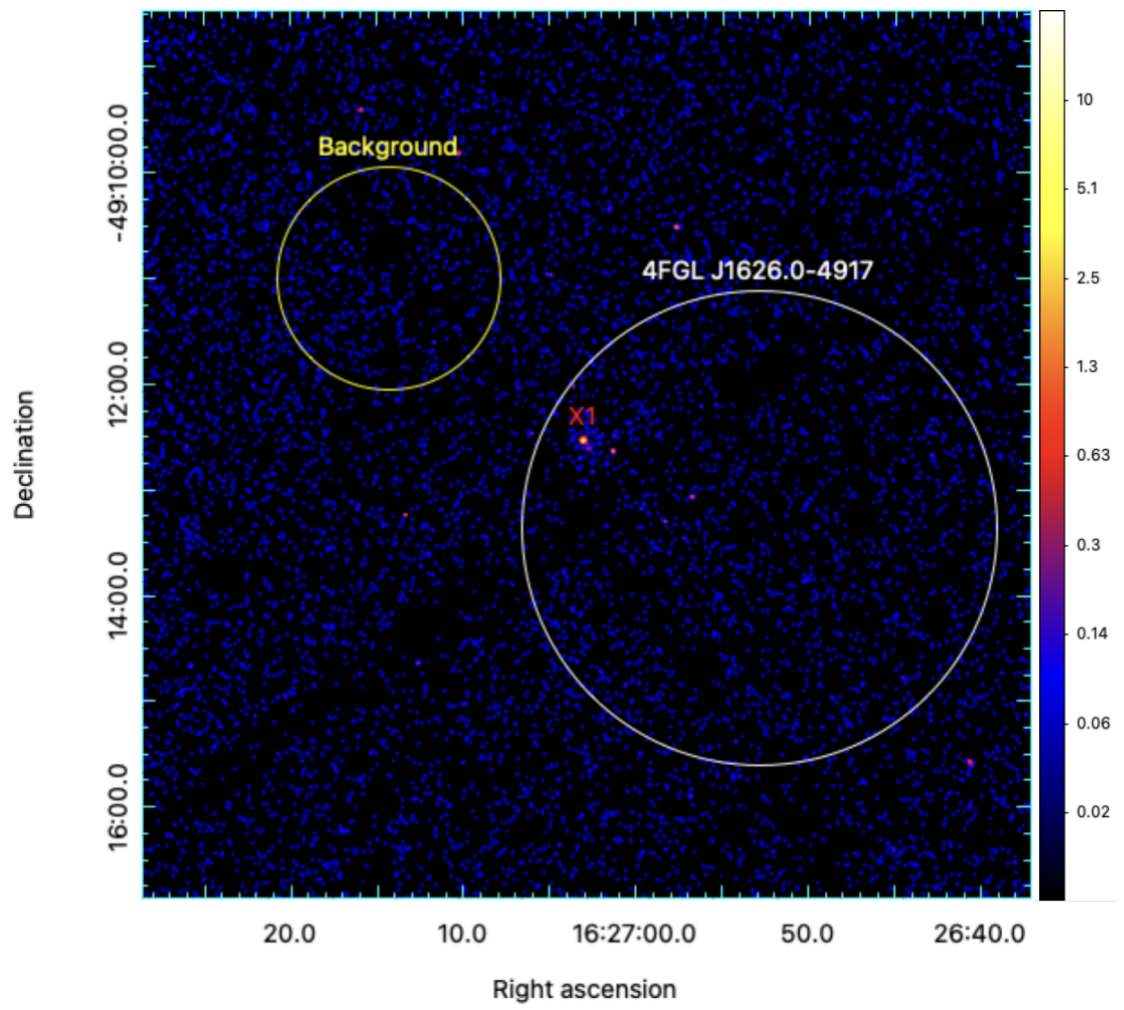}
    \caption{\textit{Chandra} X-ray sky map in energy range 0.3-8.0 keV in equatorial coordinate system. X1 (R.A.=16:27:03.0307, Dec.=-49.12.32.386) is marked in the map. The background region used for extracting spectrum is labeled as green circle. The white circle shows the location of 4FGL~J1626.0$-$4917.}
    \label{fig:j1626-chandra-map}
\end{figure}

\begin{figure}
    \centering
    \includegraphics[width=0.65\textwidth]{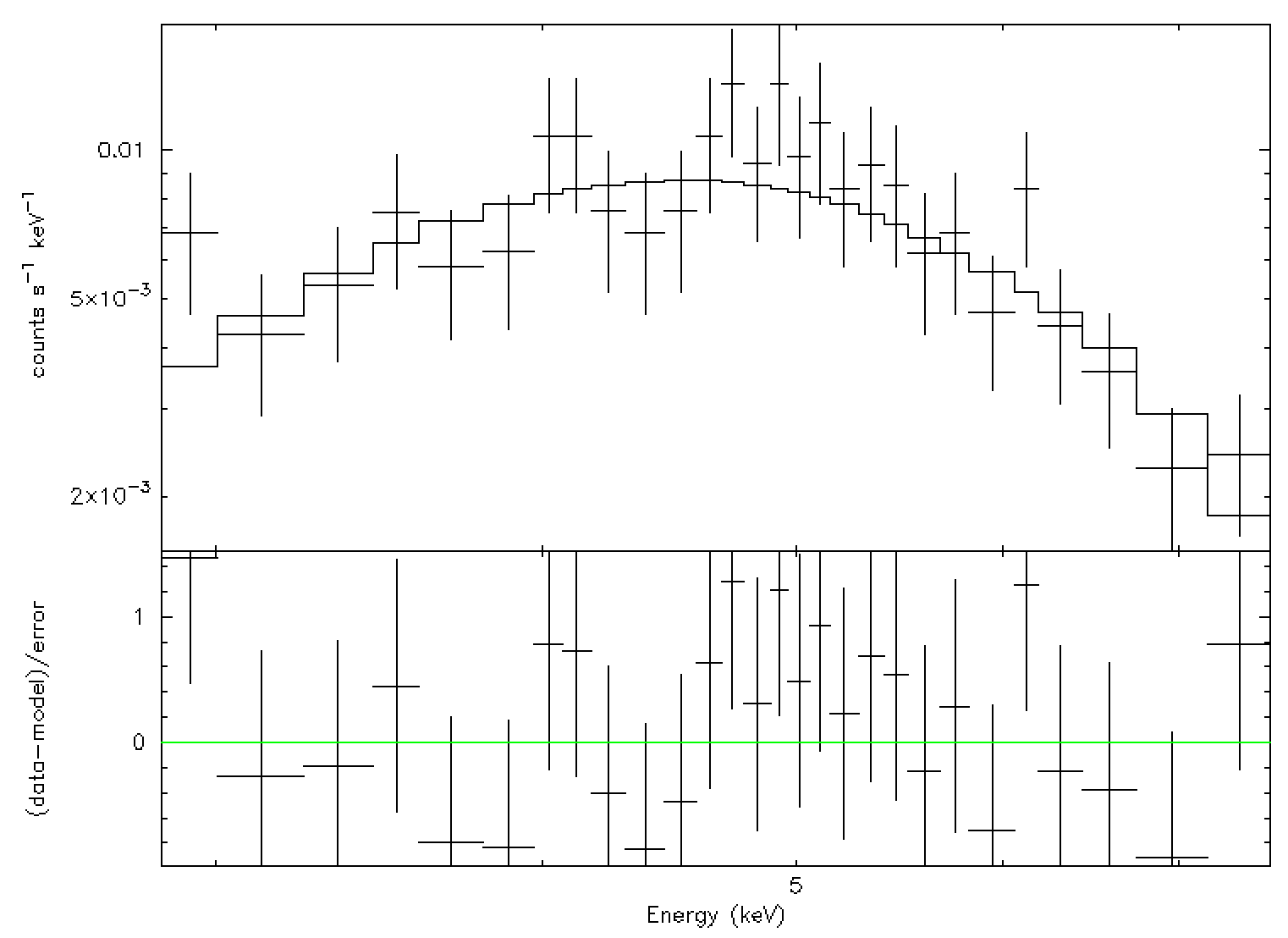}
    \caption{X-ray spectrum in the energy range of 0.3-8.0 keV with a power-law model.}
    \label{fig:j1626-chandra-spec}
\end{figure}

Figure~\ref{fig:j1626-chandra-map} shows the exposure-corrected, broadband (0.3–8.0 keV) \textit{Chandra} image and spectrum. We named the X-ray point source detected in this region as X1 (R.A.=16:27:03.0307, Dec.=-49.12.32.386). We try to unveil the nature of this X-ray source by testing different spectral models. A photoelectric absorption component \texttt{phabs} was considered for all models. Power-law, balck body, collisionless ionization equilibrium and non-equilibrium ionization models were selected to fit the spectrum respectively. The best-fit parameters are shown in Table~\ref{tab:fit}. We provide the power-law spectrum including data and model in Figure~\ref{fig:j1626-chandra-spec}.  

\begin{table}[]
    \centering
    \begin{tabular}{l|l|l}
    \hline
      Model & Best-fit Parameters & ${\chi}^2 /dof$\\\hline
       \texttt{phabs*powerlaw}  & $N_{\rm H} =(9.97 \pm 5.32) \times 10^{22}\ \rm cm^{-2}$, $\Gamma = 0.80 \pm 0.77$ & 15.42/28 \\
       \texttt{phabs*bbody}     & $N_{\rm H} =(6.80 \pm 3.81) \times 10^{22}\ \rm cm^{-2}$, $kT = 2.33 \pm 0.77\,\rm keV$  & 14.74/28\\
    \hline
    \end{tabular}
    \caption{Best-fit parameters for X-ray spectral analysis.}
    \label{tab:fit}
\end{table}

The nature of X1 can be inferred from the best-fit parameters from different models. If we considered PWN origin for the X-ray source, then the X-ray spectrum may follow a power-law model. The typical photon index $\Gamma$ for PWN in X-ray gives from 1.6 to 2.1, which is different from $0.80 \pm 0.77$ as shown in Table~\ref{tab:fit}. Similarly, once we considered the X-ray emissions from this source originated from a neutron star, then a \texttt{phabs*bbody} model should be applied. The typical temperature $kT$ of a neutron star is around 0.1 to 0.3 keV. Even for local shock acceleration, $kT$ is around 0.5 to 1 keV, which is lower than the best-fit temperature for this source. However, many neutron stars exhibit dominant non-thermal magnetospheric emission \citep{Li2008,DeGrandis2022}. In addition, the point-like source X1 might also be an AGN or post-main-sequence bluish stars, which were revealed by several X-ray observations \citep{Pizzolato2000}.

We also considered potential extended X-ray emissions in 4FGL~J1626.0$-$4917. The net count rate of 4FGL~J1626.0$-$4917 region in Figure~\ref{fig:j1626-chandra-map} is 3.41 cts/s. Notice that the net count rate of X1 is 3.35 cts/s. It means the count rate of X1 occupies more than 98\% that of 4FGL~J1626.0$-$4917. No extend X-ray emissions were found.

\section{The Origin of 4FGL~J1626.0$-$4917}\label{sec:discus}

\subsection{Stellar Cluster NGC~6134} \label{subsec:cluster}

We first consider stellar cluster NGC~6134 (R.A.=246.95, Dec=-49.16), which is an intermediate-age open cluster \citep{Tarricq2021}, as the origin of \g\ emission from 4FGL~J1626.0$-$4917. As a stellar cluster, it is located at low Galactic latitude ($\left| b \right| \textless 5^{\circ}$), which make it a possible object interacting with interstellar gas. It has a distance of $\sim 700 \ \rm pc$, and an age about 0.7 Gyr \citep{Ahumada2013}. The middle-age of NGC~6134 is suitable to study the transition of cosmic rays (CRs) from strong (young stellar cluster) to weak (old stellar cluster) stellar wind acceleration. It locates near 4FGL~J1626.0$-$4917 and HESS~J1626$-$490, which provide opportunity to study CRs via \g\ . 

Stellar clusters, which embedded in molecular clouds, are unique laboratories to understand the
interstellar medium (ISM) properties as well as the nature of interaction between young stars and ISM. They are considered as important contributor to Galactic CR \citep{Morlino2021,Blasi2023,Menchiari2024,Peron2024a,Blasi2025}. \g\ emissions have been detected from known stellar clusters, such as Westerlund 1 \citep{Aharonian2022} and Westerlund 2 \citep{Yang2018}. Also, stellar clusters are thought to be potential counterparts of \g\ unassociated sources \citep{Peron2024b}. The traditional explanation refers diffusive shock acceleration in SNRs as the principal CR production mechanism \citep{Caprioli2011,Tang2015}. Recent studies of the Cygnus cocoon and Westerlund 1 demonstrate the ambiguity inherent in distinguishing between leptonic and hadronic origins for high-energy emissions. The mounting \g\ detections coincident with stellar clusters and the emerging spectral and morphological characterizations of these sources provide circumstantial evidence supporting a contribution to the high-energy CR population, though this remains to be conclusively validated.

However, NGC~6134 is hard to be considered as the origin of 4FGL~J1626.0$-$4917. Massive OB stars are hard to be expected to survive at 0.7 Gyr. Furthermore, the average power of massive stars in a cluster drops below $10^{32}\ \rm erg/s$ for per star after 30 Myr, as pointed out by \cite{Vieu2023}. An old cluster such as NGC~6134 is not expected to be an efficient source of CRs. The energy injection rate into CRs by stellar winds is given by $\dot{E}_{\rm CR} = \eta_{\rm CR} L_{\rm w}$, where $\eta_{\rm CR}$ reflects the fraction of the wind kinetic energy which goes into accelerating CR protons. If we adopted the value $\eta_{\rm CR} = 0.1$, which is typical for supernova shock \citep{Vink2012}, the $\dot{E}_{\rm CR}$ is not high enough to maintain the CR energy.

\subsection{SNR~G335.2$+$0.1} \label{subsec:snr}

SNR~G335.2$+$0.1 (R.A.= 16:27:45, Dec.= -48:47:00) has been believed to be one of potential counterpart of HESS~J1626$-$490 \citep{Aharonian2008,Huang2025}. 
It has a shell-type radio morphology with a radio spectral index of $\alpha = 0.46$ \citep{Whiteoak1996,Green2019}. The up-to-date estimated distance is around 3.1 to 3.3 kpc. PSR J1627$-$4845 was found to locate in the area of SNR~G335.2$+$0.1. Since the pulsar has a characteristic age of $2.7 \times 10^6 \ \rm yr$, it is hard to be the product of SNR~G335.2$+$0.1 \citep{Kaspi1996}. The \g\ emission was reported by the First \lat\ SNR Catalog \citep{Acero2016}.

SNRs are important sources of accelerating particles to relativistic velocity via diffusive shock acceleration \citep{Schure2013}. Hadronic scenario has been always discussed for \lat\ SNRs \citep[e.g.][]{Fang2013,Lemoine-Goumard2025}. Extended \g\ emissions were detected in several SNRs \citep[e.g.][]{Xing2016,Ackermann2018,Abdollahi2024}. Therefore, it is meaningful to consider SNRs as \g\ sources \citep{Ajello2017}.

Section~\ref{sec:gas} and Figure~\ref{fig:sed} introduce the pp interaction of fit \g\ spectrum. The index of proton $\alpha_{\rm p} =2.32 \pm 0.14$ is a typical value for hadronic-dominated SNRs. However, \cite{Huang2025} found the \g\ extension of SNR~G335.2$+$0.1 is about $0.24^{\circ}$ in 0.2-500 GeV with a significance of $13.5\ \sigma$, which does not cover the region of 4FGL~J1626.0$-$4917. 

\subsection{Leptonic Scenario}

We tested the potential leptonic scenario which reflects the \g\ generated via the inverse Compton (IC) scattering of electrons off the seed photons around the source. For the photon field of the IC calculations, we considered the CMB radiation field, optical to UV radiation field from the star light, and the dust infrared radiation field. We calculated the IC spectrum using Naima as we did for hadronic scenario. The IC spectrum calculation is based on the formalism in \cite{Kafexhiu2014}. We assumed a exponential cutoff power-law distribution of the relativistic electrons: $f_{\rm e}(E) = A_{\rm e} (E/E_0)^{-\alpha_{\rm e}}\exp (- (E / E_{\rm c})$. The best-fit photon index of electron distribution is $\alpha_{\rm e} = 2.01 \pm 0.02$. Considering the distance of 1.8 kpc \citep{Eger2011}, we obtain the total energy above 1 GeV: $W_{\rm e} = 6.21 \times 10^{49}\ \rm erg$. We should emphasize that we can not rule out the leptonic origin.

\section{Conclusion} \label{sec:conclu}

In this paper, we report the extended \g\ emissions towards 4FGL~J1626.0$-$4917 using 17 yr \lat\ data. Our findings provide new insights into the role of extended unidentified \g\ sources in the production and propagation of CRs in the Galaxy. We find that although NGC~6134 is spatially associated with 4FGL~J1626.0$-$4917, this stellar cluster is hard to explain the \g\ emission from 4FGL~J1626.0$-$4917 according to OB stars life and wind power. The extended \g\ emissions towards 4FGL~J1626.0$-$4917 have angular extension of $0.28^{\circ}$ which cen be modelled by a Gaussian disk model. The \g\ spectrum follows a PL distribution with a photon index of $2.73 \pm 0.12$. 

The mass and number density of gas content around the region were calculated to be $1.1 \times 10^3\ \rm M_{\odot}$ and $3193\ \rm cm^{-3}$. Our study suggests that the observed \g\ emissions are probably dominated by hadronic origin, which are related to the interaction between CR protons from 4FGL~J1626.0$-$4917 and the ambient medium. However, leptonic origin can not be fully excluded.

We also investigated the potential X-ray counterpart of 4FGL~J1626.0$-$4917 using \textit{Chandra} X-ray Observatory data. An X-ray point source were found to be coincident with 4FGL~J1626.0$-$4917 spatially.

In summary, while the stellar cluster and supernova remnant are ruled out as direct counterparts, the observed morphology, spectrum, and gaseous environment collectively point towards a hadronic origin for 4FGL~J1626.0$-$4917. The source likely represents a site of ongoing CR interaction, possibly related to a population of unresolved accelerators or a faint, previously unidentified SNR/PWN. Deeper observations in the radio, X-ray, and very-high-energy \g\ regimes, particularly with instruments like the Cherenkov Telescope Array, are crucial to definitively reveal the nature of this intriguing extended source.

\begin{acknowledgements}

Scientific results from data presented in this publication are obtained from HEASARC. Ziwei Ou is supported by the National Natural Science Foundation of China (NSFC, No. 12393853). Jie Wang is supported by the National Natural Science Foundation of China (No. 12573113); the Tianshan Talent Training Program (No. 2023TSYCCX0112); the Tianshan Innovation Team Plan of Xinjiang Uygur Autonomous Region (No. 2025D14014).

\end{acknowledgements}

\bibliographystyle{raa}
\bibliography{bibtex}

\begin{thebibliography}{61}
\providecommand\natexlab[1]{#1}
\providecommand\JournalTitle[1]{#1}

\bibitem[{Abdo} {et~al.}(2010)]{Abdo2010b}
{Abdo}, A.~A., {Ackermann}, M., {Ajello}, M., {et~al.} 2010, \apjs, 188, 405

\bibitem[{Abdollahi} {et~al.}(2020)]{Abdollahi2020}
{Abdollahi}, S., {Acero}, F., {Ackermann}, M., {et~al.} 2020, \apjs, 247, 33

\bibitem[{Abdollahi} {et~al.}(2024)]{Abdollahi2024}
{Abdollahi}, S., {Acero}, F., {Acharyya}, A., {et~al.} 2024, arXiv e-prints, arXiv:2411.07162

\bibitem[{Acero} {et~al.}(2013)]{Acero2013}
{Acero}, F., {Donato}, D., {Ojha}, R., {et~al.} 2013, \apj, 779, 133

\bibitem[{Acero} {et~al.}(2016)]{Acero2016}
{Acero}, F., {Ackermann}, M., {Ajello}, M., {et~al.} 2016, \apjs, 224, 8

\bibitem[{Ackermann} {et~al.}(2018)]{Ackermann2018}
{Ackermann}, M., {Ajello}, M., {Baldini}, L., {et~al.} 2018, \apjs, 237, 32

\bibitem[{Aharonian} {et~al.}(2008)]{Aharonian2008}
{Aharonian}, F., {Akhperjanian}, A.~G., {Barres de Almeida}, U., {et~al.} 2008, \aap, 477, 353

\bibitem[{Aharonian} {et~al.}(2022)]{Aharonian2022}
{Aharonian}, F., {Ashkar}, H., {Backes}, M., {et~al.} 2022, \aap, 666, A124

\bibitem[{Ahumada} {et~al.}(2013)]{Ahumada2013}
{Ahumada}, A.~V., {Cignoni}, M., {Bragaglia}, A., {et~al.} 2013, \mnras, 430, 221

\bibitem[{Ajello} {et~al.}(2017)]{Ajello2017}
{Ajello}, M., {Atwood}, W.~B., {Baldini}, L., {et~al.} 2017, \apjs, 232, 18

\bibitem[{Araya} \& {{\'A}lvarez-Quesada}(2024)]{Araya2024}
{Araya}, M., \& {{\'A}lvarez-Quesada}, J.~A. 2024, \mnras, 527, 8006

\bibitem[{Balbo} {et~al.}(2010)]{Balbo2010}
{Balbo}, M., {Saouter}, P., {Walter}, R., {et~al.} 2010, \aap, 520, A111

\bibitem[{Blasi}(2025)]{Blasi2025}
{Blasi}, P. 2025, \aap, 694, A244

\bibitem[{Blasi} \& {Morlino}(2023)]{Blasi2023}
{Blasi}, P., \& {Morlino}, G. 2023, \mnras, 523, 4015

\bibitem[{Bolatto} {et~al.}(2013)]{Bolatto2013}
{Bolatto}, A.~D., {Wolfire}, M., \& {Leroy}, A.~K. 2013, \araa, 51, 207

\bibitem[{Caprioli} {et~al.}(2011)]{Caprioli2011}
{Caprioli}, D., {Blasi}, P., \& {Amato}, E. 2011, Astroparticle Physics, 34, 447

\bibitem[{Dame} {et~al.}(2001)]{Dame2001}
{Dame}, T.~M., {Hartmann}, D., \& {Thaddeus}, P. 2001, \apj, 547, 792

\bibitem[{De Grandis} {et~al.}(2022)]{DeGrandis2022}
{De Grandis}, D., {Rigoselli}, M., {Mereghetti}, S., {et~al.} 2022, \mnras, 516, 4932

\bibitem[{Di Mauro} {et~al.}(2019)]{DiMauro2019}
{Di Mauro}, M., {Manconi}, S., \& {Donato}, F. 2019, \prd, 100, 123015

\bibitem[{Ding} {et~al.}(2023)]{Ding2023}
{Ding}, J., {Huang}, Y., {Li}, X.-D., {et~al.} 2023, \mnras, 523, 4120

\bibitem[{Eger} {et~al.}(2011)]{Eger2011}
{Eger}, P., {Rowell}, G., {Kawamura}, A., {et~al.} 2011, \aap, 526, A82

\bibitem[{Fang} \& {Zhang}(2013)]{Fang2013}
{Fang}, J., \& {Zhang}, L. 2013, \na, 18, 35

\bibitem[{Finkbeiner}(2003)]{Finkbeiner2003}
{Finkbeiner}, D.~P. 2003, \apjs, 146, 407

\bibitem[{Ge} {et~al.}(2022)]{Ge2022}
{Ge}, T.-T., {Sun}, X.-N., {Yang}, R.-Z., {Liang}, Y.-F., \& {Liang}, E.-W. 2022, \mnras, 517, 5121

\bibitem[{Green}(2019)]{Green2019}
{Green}, D.~A. 2019, Journal of Astrophysics and Astronomy, 40, 36

\bibitem[{H.~E.~S.~S. Collaboration} {et~al.}(2018)]{Abdalla2018}
{H.~E.~S.~S. Collaboration}, {Abdalla}, H., {Abramowski}, A., {et~al.} 2018, \aap, 612, A1

\bibitem[{HI4PI Collaboration} {et~al.}(2016)]{HI4PI2016}
{HI4PI Collaboration}, {Ben Bekhti}, N., {Fl{\"o}er}, L., {et~al.} 2016, \aap, 594, A116

\bibitem[{Huang} {et~al.}(2025)]{Huang2025}
{Huang}, C., {Zhang}, X., {Chen}, Y., {et~al.} 2025, arXiv e-prints, arXiv:2507.08709

\bibitem[{Hui} {et~al.}(2020)]{Hui2020}
{Hui}, C.~Y., {Lee}, J., {Li}, K.~L., {et~al.} 2020, \mnras, 495, 1093

\bibitem[{Kafexhiu} {et~al.}(2014)]{Kafexhiu2014}
{Kafexhiu}, E., {Aharonian}, F., {Taylor}, A.~M., \& {Vila}, G.~S. 2014, \prd, 90, 123014

\bibitem[{Kaspi} {et~al.}(1996)]{Kaspi1996}
{Kaspi}, V.~M., {Manchester}, R.~N., {Johnston}, S., {Lyne}, A.~G., \& {D'Amico}, N. 1996, \aj, 111, 2028

\bibitem[{Lemoine-Goumard} {et~al.}(2025)]{Lemoine-Goumard2025}
{Lemoine-Goumard}, M., {Acero}, F., {Ballet}, J., \& {Miceli}, M. 2025, \aap, 693, A193

\bibitem[{Li} {et~al.}(2008)]{Li2008}
{Li}, X.-H., {Lu}, F.-J., \& {Li}, Z. 2008, \apj, 682, 1166

\bibitem[{Menchiari} {et~al.}(2024)]{Menchiari2024}
{Menchiari}, S., {Morlino}, G., {Amato}, E., {Bucciantini}, N., \& {Beltr{\'a}n}, M.~T. 2024, \aap, 686, A242

\bibitem[{Morlino} {et~al.}(2021)]{Morlino2021}
{Morlino}, G., {Blasi}, P., {Peretti}, E., \& {Cristofari}, P. 2021, \mnras, 504, 6096

\bibitem[{Nolan} {et~al.}(2012)]{Nolan2012}
{Nolan}, P.~L., {Abdo}, A.~A., {Ackermann}, M., {et~al.} 2012, \apjs, 199, 31

\bibitem[{Orlando} {et~al.}(2022)]{Orlando2022}
{Orlando}, E., {Rasmussen}, M., \& {Strong}, A.~W. 2022, in 37th International Cosmic Ray Conference, 662

\bibitem[{Ou}(2025)]{Ou2025}
{Ou}, Z. 2025, Astronomical Techniques and Instruments, 2, 44

\bibitem[{Paiano} {et~al.}(2017)]{Paiano2017}
{Paiano}, S., {Franceschini}, A., \& {Stamerra}, A. 2017, \mnras, 468, 4902

\bibitem[{Peron} {et~al.}(2024{\natexlab{a}})]{Peron2024a}
{Peron}, G., {Casanova}, S., {Gabici}, S., {Baghmanyan}, V., \& {Aharonian}, F. 2024{\natexlab{a}}, Nature Astronomy, 8, 530

\bibitem[{Peron} {et~al.}(2024{\natexlab{b}})]{Peron2024b}
{Peron}, G., {Morlino}, G., {Gabici}, S., {et~al.} 2024{\natexlab{b}}, \apjl, 972, L22

\bibitem[{Pizzolato} {et~al.}(2000)]{Pizzolato2000}
{Pizzolato}, N., {Maggio}, A., \& {Sciortino}, S. 2000, \aap, 361, 614

\bibitem[{Planck Collaboration} {et~al.}(2016)]{Planck2016}
{Planck Collaboration}, {Adam}, R., {Ade}, P.~A.~R., {et~al.} 2016, \aap, 594, A10

\bibitem[{Principe} {et~al.}(2020)]{Principe2020}
{Principe}, G., {Mitchell}, A.~M.~W., {Caroff}, S., {et~al.} 2020, \aap, 640, A76

\bibitem[{Sato} {et~al.}(2021)]{Sato2021}
{Sato}, S., {Kataoka}, J., {Ito}, S., {et~al.} 2021, \apj, 913, 83

\bibitem[{Saz Parkinson} {et~al.}(2016)]{SazParkinson2016}
{Saz Parkinson}, P.~M., {Xu}, H., {Yu}, P.~L.~H., {et~al.} 2016, \apj, 820, 8

\bibitem[{Schure} \& {Bell}(2013)]{Schure2013}
{Schure}, K.~M., \& {Bell}, A.~R. 2013, \mnras, 435, 1174

\bibitem[{Shi} {et~al.}(2025)]{Shi2025}
{Shi}, Y., {Cui}, Y., \& {Yang}, L. 2025, \apj, 984, 199

\bibitem[{Smith} {et~al.}(2023)]{Smith2023}
{Smith}, D.~A., {Abdollahi}, S., {Ajello}, M., {et~al.} 2023, \apj, 958, 191

\bibitem[{Tang} \& {Chevalier}(2015)]{Tang2015}
{Tang}, X., \& {Chevalier}, R.~A. 2015, \apj, 800, 103

\bibitem[{Tarricq} {et~al.}(2021)]{Tarricq2021}
{Tarricq}, Y., {Soubiran}, C., {Casamiquela}, L., {et~al.} 2021, \aap, 647, A19

\bibitem[{Tibaldo} {et~al.}(2018)]{Tibaldo2018}
{Tibaldo}, L., {Zanin}, R., {Faggioli}, G., {et~al.} 2018, \aap, 617, A78

\bibitem[{Ulgiati} {et~al.}(2025)]{Ulgiati2025}
{Ulgiati}, A., {Paiano}, S., {Pintore}, F., {et~al.} 2025, \aap, 694, A176

\bibitem[{Vieu} \& {Reville}(2023)]{Vieu2023}
{Vieu}, T., \& {Reville}, B. 2023, \mnras, 519, 136

\bibitem[{Vink}(2012)]{Vink2012}
{Vink}, J. 2012, \aapr, 20, 49

\bibitem[{Whiteoak} \& {Green}(1996)]{Whiteoak1996}
{Whiteoak}, J.~B.~Z., \& {Green}, A.~J. 1996, \aaps, 118, 329

\bibitem[{Xiang} {et~al.}(2024)]{Xiang2024}
{Xiang}, Y., {Feng}, P., \& {Lan}, X. 2024, Research in Astronomy and Astrophysics, 24, 105004

\bibitem[{Xin} {et~al.}(2019)]{Xin2019}
{Xin}, Y., {Zeng}, H., {Liu}, S., {Fan}, Y., \& {Wei}, D. 2019, \apj, 885, 162

\bibitem[{Xing} {et~al.}(2016)]{Xing2016}
{Xing}, Y., {Wang}, Z., {Zhang}, X., \& {Chen}, Y. 2016, \apj, 823, 44

\bibitem[{Yang} {et~al.}(2018)]{Yang2018}
{Yang}, R.-z., {de O{\~n}a Wilhelmi}, E., \& {Aharonian}, F. 2018, \aap, 611, A77

\bibitem[{Zabalza}(2015)]{Zabalza2015}
{Zabalza}, V. 2015, in International Cosmic Ray Conference, Vol.~34, 34th International Cosmic Ray Conference (ICRC2015), 922

\end{thebibliography}

\label{lastpage}

\end{document}